\documentclass[12pt]{article}

\usepackage{amsmath,amssymb,graphicx}

\newcommand{\dd}{\partial}
\newcommand{\de}{\delta}
\newcommand{\m}{\mu}
\newcommand{\n}{\nu}
\newcommand{\ls}{\left(}
\newcommand{\rs}{\right)}
\newcommand{\al}{\alpha}
\newcommand{\be}{\beta}
\newcommand{\ff}{\varphi}
\newcommand{\ka}{\varkappa}
\newcommand{\p}{\bot}
\newcommand{\La}{\Lambda}
\newcommand{\ta}{\tau}
\newcommand{\ro}{\rho}
\newcommand{\ep}{\varepsilon}
\newcommand{\te}{\theta}
\newcommand{\ga}{\gamma}
\newcommand{\sh}{\sinh}
\newcommand{\ch}{\cosh}

\newcommand{\disn}[2]{$$\displaylines{\refstepcounter{equation}%
            \label{#1}\hskip 1em minus 1em #2\hfilneg}$$}
\newcommand{\nom}{\hfil\hskip 1em minus 1em (\theequation)}
\newcommand{\no}{\hfil \hskip 1em minus 1em\phantom{(\theequation)}%
            \hfilneg\cr\hfilneg\hskip 1em minus 1em\hfil}

\newcommand{\st}{\\}
\makeatletter
\long\def\@makecaption#1#2{%
   \vskip 10\p@
   \setbox\@tempboxa\hbox{#1. #2}%
   \ifdim \wd\@tempboxa >\hsize
       #1. #2\par
     \else
       \hbox to\hsize{\hfil\box\@tempboxa\hfil}%
   \fi}
\makeatother

\begin{document}

\title{From the Embedding Theory to General Relativity\\ in a result of inflation}

\author{S.A.~Paston\thanks{E-mail: paston@pobox.spbu.ru},
A.A.~Sheykin\thanks{E-mail: anton.shejkin@gmail.com}\\
{\it Saint Petersburg State University, St.-Petersburg, Russia}
}
\date{\vskip 15mm}
\maketitle

\begin{abstract}
We study the embedding theory being a formulation of the gravitation theory where the independent variable is the embedding function for the four-dimensional space-time in a flat ambient space.
We do not impose additional constraints which are usually used to remove from the theory the extra solutions not being the solutions of Einstein equations.
In order to show the possibility of automatic removal of these extra solutions we analyze the equations of the theory, assuming an inflation period during the expansion of the Universe.
In the framework of FRW symmetry we study the initial conditions for the inflation, and we show that after its termination the Einstein equations begin to satisfy with a very high precision.
The properties of the theory equations allow us to suppose with confidence
that the Einstein equations will satisfy with enough precision
out of the framework of FRW symmetry as well.
Thus the embedding theory can be considered as a theory of gravity which explains observed facts without any additional modification of it and we can use this theory in a flat space when we try to develop a quantum theory of gravity.
\end{abstract}


\newpage
\section{Introduction\label{p1}}
The embedding theory was suggested as an alternative formulation of the gravitation theory
by Regge and Teitelboim in 1975 in the article \cite{regge}.
They suggested to consider our space-time to be a four-dimensional
surface in a Minkowski space of higher dimensions $R^{1,N-1}$
(where $N=10$ appears to be the most natural in general case).
As a result the dynamics of three-dimensional space is described
similarly to the dynamics of a point particle to which corresponds a world line,
and to a string dynamics to which corresponds a two-dimensional surface
in the Minkowski space.
This analogy is the main motivation of the approach suggested by Regge and Teitelboim.
In the framework of this approach
the independent variable describing the gravity is the embedding function $y^a(x)$
of this surface into the ambient space,
while the metrics becomes induced and it is expressed by the formula
 \disn{n1}{
g_{\m\n}=(\dd_\m y^a)\eta_{ab}(\dd_\n y^b).
\nom}
Here and below $\m,\n,\ldots=0,1,2,3$; $a,b,\ldots=0,1,\ldots,N-1$, and $\eta_{ab}$
is the constant metrics of the ambient space $R^{1,N-1}$.

Note that the embedding theory strongly differs from the theory
with extra dimensions which became popular in recent decades
(see for example the overview \cite{barvin})
by the fact that in the embedding theory the ambient space is flat,
i.e. there is no gravity in it.
Note however the work \cite{davgur06} where these two approaches are considered together.

The approach to the gravity based on the theory of a surface in a flat ambient space
might be more convenient than a standard approach when we try to develop a quantum theory of gravity.
The problems arising when we attempt to quantize the gravitation in terms of metric $g_{\m\n}$ are discussed
in the overview \cite{carlip}. Most of them are due to the fact that the quantization procedure convenient
for field theory in a flat space is used to
quantize the dynamic variables being just the geometric properties of the space.

One of the problems is the "problem of time".
Since in the framework of diffeomorphism symmetry the theory have invariance with respect to time reparametrization,
the Hamiltonian reduces to a linear combination of constraints
(see the Arnowitt-Deser-Misner formalism \cite{adm}).
Hence it vanishes while acting on the vectors of the physical space where the constraints are satisfied.
As a result, the vector describing the state of the system in a physical sub-space in the framework of Schr{\"o}dinger representation does not change with time.
Usually it is supposed to mean that the "physical" time is represented not by the $x^0$ coordinate, but by another value,
however there is no satisfactory method to build this value in General Relativity.
Since the embedding theory is formulated in a flat ambient space, the role of the "physical"
time can be played by a time-like coordinate.
Note that this problem needs a separate consideration,
since a canonical formalism for the embedding theory appears to be very complicated \cite{tapia,frtap,davkar,rojas06},
as it contains the second class constraints.

One more problem arises when we try to formulate the causality principle in gravity.
The causality principle in quantum field theory usually means the commutativity of operators related to areas separated by a space-like interval. This principle is difficult to formulate in standard gravity in terms of metric $g_{\m\n}$, because the interval between points is given by the metric which is an operator itself.  Therefore it is impossible to determine for two given points of the space-time what kind of interval separates them absolutely, independently of a specific vector of state. But if gravity is described as a four-dimensional surface in a flat ambient space,
we can try to work out a quantum field theory in this space which gives this gravity in classical limit. If it will be done, the problem of formulation of the causality principle would be solved,
since causality in the flat ambient space can be determined by standard means of quantum field theory.
In \cite{tmf2011} an example of a field theory in a flat space which describes many non-interacting four-dimensional surfaces is suggested.
Each of them satisfies the same equations as in the embedding theory and can be considered to be our space-time.

After the article \cite{regge} the idea of embedding
has been critically discussed in the work \cite{deser},
and later this idea was quite often used for description of gravity,
including its quantification, see for example the works
\cite{pavsic85let,tapia,maia89,bandos,davkar}.
A detailed analysis of motion equations in the embedding theory compared to the Einstein equations has been performed in \cite{tmf07}.
Among recent works using the idea of embedding we note \cite{faddeev,rojas09}.
An extended bibliography related to the embedding theory and similar problems can be found in \cite{tapiaob}.

One of main difficulties of the embedding theory is that if we take for the action the standard Einstein-Hilbert expression formulated in terms of the embedding function, then the arising equations of motion (the "Regge-Teitelboim equations") are more general than the Einstein equations.
In this case all the solutions of the Einstein equations satisfy the Regge-Teitelboim equations,
but the latter ones contain also some "extra" solutions.
In order to overcome this problem it has been suggested \cite{regge} to impose additional constraints ("Einstein's constraints"),
which in the absence of matter read $G_{\m\p}=0$, i.~e. they are a part of the Einstein equations (a generalization to the presence of matter is not difficult).
Here $G_{\m\n}$ is the Einstein tensor, and the symbol $\p$ denotes the direction orthogonal to the constant time surface.
It has been shown in \cite{tmf07,ijtp10} that if the Einstein's constraints are imposed at the initial instant, all the Einstein equations will be satisfied during the evolution of the system. A canonical formalism was construct describing this evolution, and a corresponding full constraint algebra was found. The theory worked out in this way is equivalent to the Einstein's General Relativity,
in particular, it is easy to follow (see ~\cite{tmf07,ijtp10}) the relation between the constraint algebra of this theory and the algebra describing the General Relativity in the Arnowitt-Deser-Misner approach.

However, as noted in \cite{deser},  artificial {\it ad hoc} imposing of additional constraints to the theory seems not quite satisfactory.
A more attractive possibility is to write down an action where the Einstein's constraints appear as a part of the Euler-Lagrange equations, as it was done in \cite{tmf07,ijtp10}. But the suggested action appears too artificial and contains explicitly a singled out direction of time, being its important defect.

The main aim of this paper is to show that the embedding theory
is equivalent to General Relativity with great accuracy in assumption of existence of
exponential expansion period (inflation) during the evolution of the Universe.
This makes it possible to consider the embedding theory
as a theory of gravity which explains observed facts without any additional modification of it.
Therefore we can use this theory in a flat space when we try to develop a quantum theory of gravity.

We reject the idea of artificial imposing of constraints and we analyze the embedding theory in its initial form.
We will make and try to validate for the embedding theory the assumption that
if there has been an inflation period, then after its termination the Einstein equations begin to satisfy with a high precision.
In section~\ref{p2} we present some general ideas about this hypothesis,
and in section~\ref{p3}, assuming the Friedmann-Robertson-Walker (FRW) symmetry,
we exactly analyze the possible scenarios of the Universe evolution in the framework of the embedding theory,
if during the first classical epoch the Universe was filled with matter imitating the $\La$-term.
Note that the embedding theory in the FRW symmetry framework
has been studied in a similar way in the works \cite{davids97,davids01},
but there has not been analyzed how much the deviations from the exact solutions of the Einstein equations decrease during the inflation.

We will show
that three evolution scenarios are possible depending on the initial conditions at the instant when classical equations
of motion start satisfying (we can suppose an unknown quantum gravity before).
In one of these scenarios the expansion is succeeded by compression and the Universe
does not probably reach macroscopic dimensions.
In another one the expansion is infinite and follows not an exponential,
but a power law without acceleration (in spite of the presence of matter imitating the $\La$-term).
In the third one the Universe expands exponentially.
Hence, in two of the three scenarios there is no inflation,
while the inflation is needed to explain the astrophysical observations.

In the third scenario,
if we suppose that before the beginning of the inflation
the deviations from the exact satisfying of Einstein equations
do not differ too much from the unity
(the motivation of this assumption is given in the beginning of section~\ref{p4}),
then at the end of the exponential expansion these deviations appear to be very small,
of the order of $10^{-44}$.
The discussion of the Universe expansion in the framework of the embedding theory based
on the results obtained in the framework of FRW symmetry is given in section~\ref{p4}.

Since during the inflation the FRW approximation begins to work very well,
we can conclude, in particular, that the Einstein's constraints
(which were suggested in \cite{regge} to be imposed in an artificial way),
begin to satisfy quite automatically with high precision in a result of the inflation.
As already said above, the exact imposition of Einstein's constraints
at some moment of time results in satisfying the Einstein equations after this moment.
For this reason satisfying the Einstein's constraints
with the precision of the order of $10^{-44}$
allows to suppose with confidence that
the deviations from the exact solutions of the Einstein equations after the end of inflation
remain very small even out of the FRW symmetry framework.

\section{The idea of appearance of the Einstein's dynamics\st as a result of inflation\label{p2}}
Consider the equations of motion in the embedding theory. Following \cite{regge}, we take for the action the standard expression
 \disn{1}{
S=\int d^4x\sqrt{-g}\ls-\frac{1}{2\ka} R+{\cal L}_m\rs,
\nom}
where the metrics is expressed by the formula (\ref{n1}) via the embedding function $y^a(x)$ being an independent variable,
${\cal L}_m$ is the contribution of fields of matter
and $\ka$ is the gravitation constant.
The Regge-Teitelboim equations being the equations of motion corresponding to the taken action read:
 \disn{3}{
D_\m\ls\ls G^{\m\n}-\ka\, T^{\m\n}\rs\dd_\n y^b\rs=0,
\nom}
where $D_\m$ is the covariant derivative,
and $T^{\m\n}$ is the energy-momentum tensor of matter.
Obviously, any solution of the Einstein's equations is also a solution of these equations.
Note that it follows from (\ref{3}) (see~\cite{tmf07}) one more form of the motion equations
 \disn{2}{
\ls G^{\m\n}-\ka\, T^{\m\n}\rs b^b_{\m\n}=0
\nom}
and, taking into account the Bianchi identities, the relation
 \disn{4}{
D_\m T^{\m\n}=0,
\nom}
which becomes an identity if the equations of motions for matter are satisfied.
Here $b^b_{\m\n}=D_\m\dd_\n y^b$ is the second fundamental form of the surface.

Denote
 \disn{5}{
\ta^{\m\n}\equiv \frac{1}{\ka}G^{\m\n}-T^{\m\n}.
\nom}
Using it we can suppose that the Einstein equations
 \disn{6}{
G^{\m\n}=\ka\ls T^{\m\n}+\ta^{\m\n}\rs
\nom}
satisfy with an additional term to the energy-momentum tensor of matter
(this interpretation was used in \cite{davkar}, and
$\ta^{\m\n}$ can be named the energy-momentum tensor of Dark Matter).

The resulting theory has the form of General Relativity with additional "$\ta$-matter" giving the contribution $\ta^{\m\n}$ to the full energy-momentum tensor.
The Regge-Teitelboim equations (\ref{3}) appear from this point of view to be some limitations for the $\ta^{\m\n}$ under the form
 \disn{8}{
D_\m\ls\ta^{\m\n}\,\dd_\n y^b\rs=0.
\nom}
We note again that two relations follow from (\ref{8}):
 \disn{7}{
\ta^{\m\n} b^b_{\m\n}=0
\nom}
and
 \disn{9}{
D_\m\ta^{\m\n}=0.
\nom}

We analyze now the equation (\ref{8}). Under the covariant derivative we have a vector
with respect to a diffeomorphism group (in contrast to the equation (\ref{4})), because the free index $a$ does not relate to this group.
This is why the equation can be rewritten as a usual divergence equal to zero
 \disn{10}{
\dd_\m\ls\sqrt{-g}\,\ta^{\m\n}\,\dd_\n y^b\rs=0,
\nom}
i.~e. under the form of the continuity equation for a $N$-component density of current:
 \disn{11}{
\dd_\m j^\m_b=0,\qquad
j^\m_b=\sqrt{-g}\,\ta^{\m\n}\,\dd_\n y_b.
\nom}
It means, in particular, the existence of $N$ conserved quantities.

The satisfied continuity equation for the density of current $j^\m_b$ allows to suppose that as more the Universe expands, this quantity will decrease inversely to a power of the scaling factor, because the corresponding conserved quantities distribute over the increasing volume.
At least, if the quantity $j^\m_b$ is a time-like vector (by the index $\m$) distributed in a sufficiently homogeneous manner, it will surely decrease in a indicated manner during the Universe expansion.
Since the quantity $\ta^{\m\n}$ can be expressed via $j^\m_b$ by the formula
\disn{12}{
\ta^{\m\n}=\frac{1}{\sqrt{-g}}\,j^\m_b\, g^{\n\al}\dd_\al y^b,
\nom}
the decrease of $j^\m_b$ will also lead to its decrease.
Assuming FRW symmetry to be satisfied, one can work out more strict conclusions concerning the decrease of the quantity $\ta^{\m\n}$, the corresponding analysis will be performed below in section~\ref{p3}.

Note that, as it is well known, the equation (\ref{4}) (in contrast to the equation (\ref{8})) does not always ensure the decrease of the components of $T^{\m\n}$ at the Universe expansion.
For example, at
 \disn{12.1}{
T^{\m\n}=\La\, g^{\m\n},
\nom}
which corresponds to the presence of the $\La$-term (or of matter which imitates it), the equation (\ref{4}) satisfies at constant $\La$.
According to the modern knowledge
(see for example the monograph \cite{gorbrub}),
during the first classical epoch of the Universe evolution
(when the classical equation of motion became valid)
the main contribution to the energy-momentum tensor was given by matter imitating the $\La$-term, i.~e. the relation (\ref{12.1}) was approximately satisfied.
During the period when it is so the General Relativity predicts an exponential expansion Universe, i.~e. the inflation.
The hypothesis of the inflation epoch is needed to solve known problems of the  hot Big Bang theory, see~\cite{gorbrub}.

Since during the inflation the parameter $\La$ changes very slowly with the increase of $a$, the power decrease of $\ta^{\m\n}$ results in the fact that this value in (\ref{6}) becomes only a small correction to the value $T^{\m\n}$ having the form of (\ref{12.1}). It means in fact that during the inflation the usual Einstein equations begin to satisfy almost exactly.
Hence, the dynamics of the embedding theory which in general case does not necessary follow the Einstein equation (because "extra" solutions are possible) becomes the Einstein's dynamics as a result of inflation.
Note that it is not strictly proven in general case because at its validation some assumptions were made. However for the interesting case of FRW symmetry which is usually supposed at the inflation epoch, this statement can be strictly proven. This is the subject of the next section.

\section{The exact analysis in the framework\st of FRW symmetry\label{p3}}
\subsection{Deducing the equations}
We analyze now the embedding theory equations assuming that FRW symmetry takes place.
In this case the energy-momentum tensor of matter can be written under the form:
 \disn{13}{
T^{\m\n}=(\ro+p)u^\m u^\n-p\, g^{\m\n},
\nom}
where $\ro$, $p$ are the density and the pressure of matter (they depend only on time), and
$u^\m$ is the unit time-like vector orthogonal to the constant time surface.
The quantities $\ro$ and $p$ are interconnected by an equation of state of matter which will be firstly assumed to be arbitrary.

Since in the embedding theory the independent variable is the embedding function $y^a(x)$,
we assume that the four-dimensional surface described by it in the ambient space also corresponds to FRW symmetry. That means that its constant time cross sections are homogeneous and isotropic spaces from the point of view of the external geometry.
This condition is easily satisfied for closed and open FRW models if the constant time submanifolds will be a three-dimensional sphere and pseudosphere (hyperboloid) respectively.
For the embedding function for the case of the spatially flat model we will use the expression obtained with the help of a limit transition from a closed (or open) model
assuming that the surface for the spatially flat model is close to the corresponding surface for the closed (open) model at a big radius of curvature of the constant time cross section.
The explicit form of the used embedding functions for all the three FRW models, as well as the explanations why they are used can be found in the Appendix, equations (\ref{14}), (\ref{18}), (\ref{21}).
Note that in all the three cases, if FRW symmetry is satisfied exactly, the ambient space appears to be five-dimensional, i.~e. $N=5$.

The fact that the metric for all the three FRW models corresponds to the homogeneity and isotropy of the constant time surfaces leads to the conclusion that the form of the corresponding Einstein's tensor must be similar to the formula (\ref{13}).
Owing to the definition (\ref{5}) the same can be said concerning the quantity $\ta^{\m\n}$,
i.~e. in case of FRW symmetry it can be written in the form of
 \disn{23}{
\ta^{\m\n}=(\ro_\ta+p_\ta)u^\m u^\n-p_\ta\, g^{\m\n},
\nom}
where the parameters $\ro_\ta$ and $p_\ta$ depending only on the time can be called the density and the pressure
of "$\ta$-matter".
Note that since the value $\ta^{\m\n}$ was introduced in a formal way, there are no {\it a priori} limitations to the values $\ro_\ta$ and $p_\ta$ (in contrast to $\ro$ and $p$); for example, $\ro_\ta<0$ is quite possible.

Now, using the introduced denotations, rewrite the corollaries of the Regge-Teitelboim equations (\ref{3}), i.~e. the equations (\ref{6}), (\ref{4}), (\ref{9}), (\ref{10}) for the case of FRW symmetry.
The Einstein equation (\ref{6}) with additional term $\ta^{\m\n}$ can be written in the form of FRW equations containing also an additional contribution:
 \disn{24}{
\ls\frac{\dot a}{a}\rs^2=\frac{\ka}{3}\ls \ro+\ro_\ta\rs-\frac{\ep}{a^2},
\nom}
 \disn{25}{
\frac{\ddot a}{a}=-\frac{\ka}{6}\ls \ro+\ro_\ta+3(p+p_\ta)\rs.
\nom}
Here $a$ is a time-dependent scaling factor
(for closed and open models it is equal to the radius of curvature of the three-dimensional space;
its exact definition results from the expressions for the interval (\ref{17}), (\ref{20}), (\ref{22}) given in the Appendix),
the point means the time derivative
(note that $g_{00}=1$), and $\ep=1,-1,0$ for closed, open and spatially flat models respectively.

The equations (\ref{4}) and (\ref{9}) having the same form are written in a standard way:
 \disn{26}{
\dd_0(\ro\, a^3)+p\,\dd_0(a^3)=0,
\nom}
 \disn{27}{
\dd_0(\ro_\ta a^3)+p_\ta\dd_0(a^3)=0.
\nom}
As usual, if we do not consider static solutions $a=const$,
the equation (\ref{25}) follows from the equations (\ref{24}) and (\ref{26}),(\ref{27}) and can be ignored. Besides, in this case we can express the quantity $p_\ta$ from the equation (\ref{27}):
 \disn{50}{
p_\ta=-\ro_\ta-\frac{a}{3\dot a}\,\dot\ro_\ta.
\nom}

Consider the equation (\ref{10}).
When the relation (\ref{27}) is satisfied (and therefore (\ref{9}) is also satisfied) the equation (\ref{10}) becomes (\ref{7}) and therefore all its components except one are satisfied identically. It follows from the fact that the second fundamental form of the surface $b^b_{\m\n}$ is orthogonal  with respect to the index $b$ to any vector tangent to the surface, and the surface has a codimension 1, because $N=5$.
Hence we can limit our consideration to only one component of the equation (\ref{10}).
We will choose this component for each of the three FRW models in order to ensure the dependence of the corresponding component of the embedding function $y^b$ only on the time.
As seen from the formulas (\ref{14}),(\ref{18}),(\ref{21}), we must take $b=0$ in order to ensure this for the closed model,
$b=4$ for the open model,
and the difference of the components corresponding to $b=0$ and $b=1$ for the spatially flat model.
Then, using the fact that the selected component $y^b$ depends only on time, we apply the formula (\ref{23}) and we note that $u^\m=\de^\m_0$,
and we also use the expression for the determinant of the metrics $g$ and the formulas (\ref{15}),(\ref{19}). As a result we found the form of the equation (\ref{10}) in case of FRW symmetry:
 \disn{31.1}{
\dd_0 \ls a^3\ro_\ta\sqrt{\dot a^2+\ep}\rs=0.
\nom}
Note that the radicand is non-negative even for the open model (when $\ep=-1$) due to (\ref{19}).

The equation (\ref{31.1}), with the use of (\ref{24}), allows to easily obtain the algebraic relation
(it has been obtained in \cite{davids01})
 \disn{42}{
a^4\ro_\ta\sqrt{\ro+\ro_\ta}=C,
\nom}
where $C$ is the constant given by the initial conditions.
Here the radicand is still non-negative, i.~e.
 \disn{41}{
\ro_\ta\geqslant-\ro.
\nom}
If at the initial instant the condition $\ro_\ta=0$ is met, the constant $C$ is equal to zero. It follows that at any subsequent instant the relation $\ro_\ta=0$ will be met as a result of the equation (\ref{42}), and $p_\ta=0$ as a result of the relation (\ref{50}), i.~e. the Einstein equations (\ref{6}) will be met without additional term $\ta^{\m\n}$.
This fact is a particular case of the general result obtained in \cite{tmf07,ijtp10} (see Introduction).

Using the obtained equations (\ref{24}), (\ref{26}), (\ref{50}), (\ref{42}) we now analyze the process of the Universe expansion starting from the instant when the classical equations of motion begin to satisfy.
The initial data corresponding to this instant will be marked by "in":
$\ro^{\text{in}}$, $a_{\text{in}}$, $\dot a_{\text{in}}$.
The equation (\ref{24}) gives for $\ro_\ta^{\text{in}}$:
 \disn{41.1}{
\ro_\ta^{\text{in}}=-\ro^{\text{in}}+\frac{3}{\ka}\ls\frac{\dot a_\text{in}^2+\ep}{a_\text{in}^2}\rs.
\nom}
Let the initial data correspond to the general case, then, in particular, $\ro_\ta^{\text{in}}\ne 0$, $\ro_\ta^{\text{in}}+\ro^{\text{in}}\ne 0$.
This implies that in (\ref{42}) we obtain $C\ne 0$, that the sign of $\ro_\ta$ is unchanged in time and coincides with the sign of the constant $C$, and that the formula (\ref{42}) implies
\disn{47}{
a=\ls\frac{|C|}{|\ro_\ta|\sqrt{\ro+\ro_\ta}}\rs^{1/4}\!=
a_{\text{in}}\ls\frac{|\ro_\ta^{\text{in}}|\sqrt{\ro^{\text{in}}+\ro_\ta^{\text{in}}}}{|\ro_\ta|\sqrt{\ro+\ro_\ta}}\rs^{1/4}\!\!\!\!.
\nom}
We also suppose that $\dot a^{\text{in}}>0$, because otherwise the size of the Universe will decrease and the classical equations of motion will not satisfy any more.

\subsection{The epoch of matter imitating the $\La$-term}
As was already said in the end of section~\ref{p2},
we suppose that during the first classical epoch of the Universe evolution
the main contribution to the energy-momentum tensor was given by matter imitating the $\La$-term, for which the equation of state $p=-\ro$ was almost exactly satisfied.
The equation for this matter (\ref{26}) results in a constant $\ro$ (as much exact as the mentioned equation of state is satisfied).
In this case the General Relativity results unambiguously in the exponential expansion of the Universe: the inflation.
We now study the consequences of the embedding theory in this case.
To do this we have to analyze the solutions of the system of equations (\ref{24}),(\ref{47})
at $\ro=\La=const$.

If we substitute $a$ from (\ref{47}) into the right part of the equation (\ref{24}),
then we obtain the equation
\disn{51}{
\ls\frac{\dot a}{a}\rs^2=\frac{\La+\ro_\ta}{\sqrt{|C|}}\ls
\frac{\ka}{3}\sqrt{|C|}-\ep F(\ro_\ta)\rs,
\nom}
where we denoted
\disn{51.1}{
F(\ro_\ta)=\frac{|\ro_\ta|^{1/2}}{(\La+\ro_\ta)^{3/4}}.
\nom}
We now study what will happen at various initial values of $\ro_\ta^{\text{in}}$. Note that due to the condition (\ref{41}) and due to the assumption of non specific  initial data made after the formula (\ref{41.1}), the value $\ro_\ta^{\text{in}}$ is limited by the condition $\ro_\ta^{\text{in}}>-\La$.

\begin{figure}[htb]
\centerline{\includegraphics[width=0.7\textwidth]{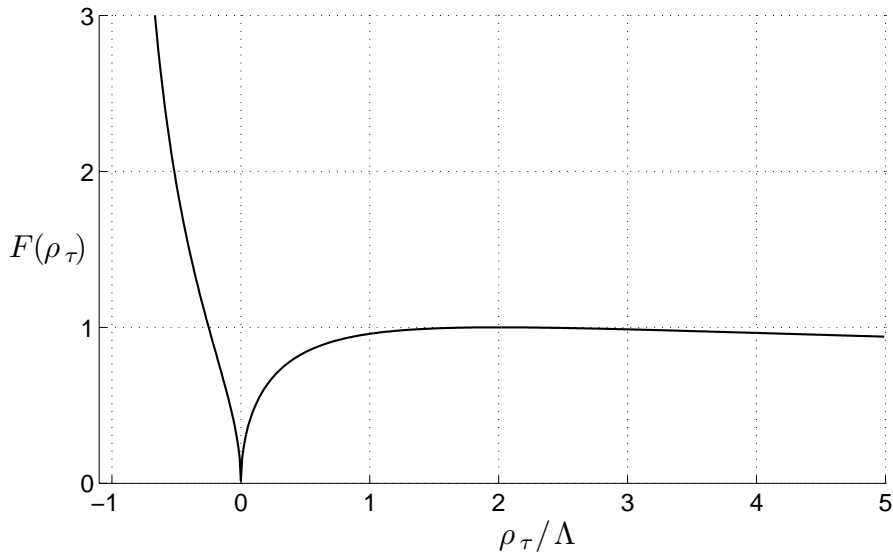}}
\caption{\label{ris2}
The graphic of the function $F(\ro_\ta)$ given by (\ref{51.1}).
The maximum at $\ro_\ta>0$ is reached at the point $\ro_\ta=2\La$.
}
\end{figure}

First consider the case when $\ro_\ta^{\text{in}}>0$, hence $\ro_\ta,C>0$.
\begin{figure}[htb]
\centerline{\includegraphics[width=0.7\textwidth]{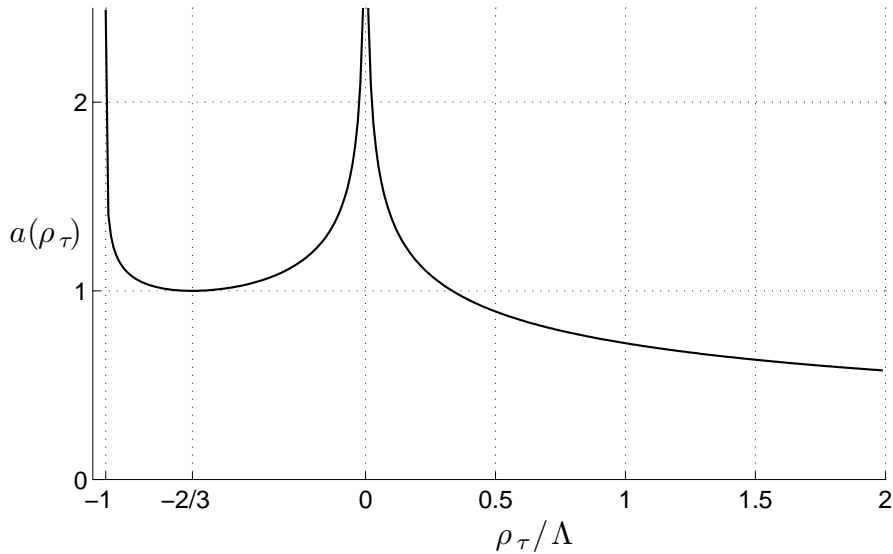}}
\caption{\label{ris1}
The dependence $a(\ro_\ta)$ given by the formula (\ref{47}).
The minimum at $\ro_\ta<0$ is reached at the point $\ro_\ta=-\frac{2}{3}\La$.
}
\end{figure}
As seen from fig.~\ref{ris1} showing the dependence $a(\ro_\ta)$ given by the formula (\ref{47}), the time increase means moving to the left of the graph, in the direction of the decrease of $\ro_\ta$ (taking into account the assumption $\dot a^{\text{in}}>0$).
Then there are two possibilities.
If $\ep=0,-1$, then the right side of the equation (\ref{51}) is positive, moreover it is distant from zero by a positive constant.
The same assertion concerns $\ep=1$, if the maximum of the function $F(\ro_\ta)$ (shown on the fig.~\ref{ris2})
in the region $\ro_\ta\ge 0$ is strictly lower than $\ka\sqrt{|C|}/3$, which is satisfied, as can be shown, if
\disn{51.2}{
\ka\,\La\, a_{\text{in}}^2\ls\dot a_{\text{in}}^2+1\rs\ls\dot a_{\text{in}}^2+1-\frac{1}{3}\ka\,\La\, a_{\text{in}}^2\rs^2
>\frac{4}{9}.
\nom}
In this case an exponential expansion Universe, i.~e. the inflation, will occur, where $\ro_\ta$ will steadily decrease and tend to zero at $a\to\infty$.
If at $\ep=1$ the inequality (\ref{51.2}) is not satisfied, then a forbidden range appears for values of $\ro_\ta$, where the right side of (\ref{51}) is negative.
In this case, if $\ro_\ta^{\text{in}}$ is lower than the values from this range, that is satisfied, as one can see, when
\disn{51.3}{
\ka\,\La\, a_{\text{in}}^2>1,
\nom}
then the inflation again takes place (taking into account the assumption $\dot a^{\text{in}}>0$).
But if $\ro_\ta^{\text{in}}$ is greater than these values, then the expanding Universe reaches only a finite value $a_{\text{max}}$, corresponding to $\dot a=0$.
For $a_{\text{max}}$ one can obtain an upper estimate
\disn{78}{
a_{\text{max}}\leqslant\frac{1}{\sqrt{\ka\,\La}}.
\nom}
The parameter $a$ reaches $a_{\text{max}}$ either asymptotically (if the relation (\ref{51.2}) becomes the equality), or in a finite time, after which the expansion is succeeded by the compression.

Now consider the case when $-\frac{2}{3}\La<\ro_\ta^{\text{in}}<0$, and hence $\ro_\ta,C<0$ (note that the function $a(\ro_\ta)$ shown on fig.~\ref{ris1} has minimum at $\ro_\ta=-\frac{2}{3}\La$).
Assuming again $\dot a^{\text{in}}>0$, we easily see on fig.~\ref{ris1}, that in this case the time increases to the right of the graph, in the sense of the decrease of $|\ro_\ta|$.
In this case, as we see on fig.~\ref{ris2}, the values of the function $F(\ro_\ta)$ will decrease, hence, for $\ep=1$ the right side of the equation (\ref{51}) will be distant from zero by its initial value.
But if $\ep=0,-1$, it will also be distant from zero by a positive constant, hence at $\ro_\ta^{\text{in}}$ from the given range the inflation will surely take place, and during the inflation the value $\ro_\ta$ will steadily decrease, approaching to zero from the negative side at $a\to\infty$.

Finally, consider the case of $-\La<\ro_\ta^{\text{in}}<-\frac{2}{3}\La$, where again  $\ro_\ta,C<0$.
Assuming again $\dot a^{\text{in}}>0$, we easily see on fig.~\ref{ris1}, that in this case the time increases to the left of the graph, in the sense of the increase of $|\ro_\ta|$.
If $\ep=1$, then, as we see on fig.~\ref{ris2}, at some value of $a_{\text{max}}$ the right side of the equation (\ref{51}) becomes equal to zero, resulting in the reversion of the expansion by the compression. In this case the value $a_{\text{max}}$ can be estimated
\disn{85}{
a_{\text{max}}\leqslant
a_{\text{in}}\frac{\ls\ka\,\La\, a_{\text{in}}^2\rs^{3/2}}{18}.
\nom}
The compression will occur up to the minimum value of $a$, corresponding to the point $\ro_\ta=-\frac{2}{3}\La$ where the function shown on fig.~\ref{ris1} takes its minimum.
Analyzing the equations in the vicinity of this point one can show that it will be reached in a finite time, where $\dot a$ will be finite and $\ddot a$ will tend to infinity.
As we see from (\ref{25}), that means that $p_\ta$ becomes infinite, i.~e. the space components of the Einstein's tensor are infinite, and the classical equations in the vicinity of this point are no more valid.

But if $\ep=0,-1$, the right side of the equation (\ref{51}) does not become equal to zero at finite $a$, but in contrast to the cases considered above leading to the inflation, it is not separated from zero by a positive constant.
We analyze now in this case the Universe expansion.
As seen from (\ref{42}), after the increase of the parameter $a$ just in several times the value $\ro_\ta$ becomes close to $-\La$, and the following asymptotic can be used
 \disn{53}{
\ro_\ta\approx -\La+\frac{C^2}{\La^2 a^8}.
\nom}
Substituting this asymptotic in (\ref{24}) we get:
 \disn{54}{
\ls\frac{\dot a}{a}\rs^2=\frac{\ka\, C^2}{3\La^2 a^8}-\frac{\ep}{a^2}.
\nom}
It follows from this equation that
for $\ep=0$ the expansion will follow the law $a\sim t^{1/4}$,
and for $\ep=-1$ it will initially follow the same law, but then (when the second term in the right side of (\ref{54}) will dominate) it will follow the law $a\sim t$.
Hence the expansion will follow a slow power law (without acceleration), and there will be no inflation.

The results obtained for all the considered cases allow together to make the following conclusions for the first classical epoch where the main role is played by matter imitating the $\La$-term. In open and spatially flat FRW models ($\ep=-1,0$) the inflation occurs (the "inflation" scenario is realized) if the condition $\ro_\ta^{\text{in}}>-\frac{2}{3}\La$ takes place, which can be written, using (\ref{41.1}), under the form of a limitation of the initial data
\disn{54.1}{
\frac{\dot a_\text{in}^2+\ep}{\ka\,\La\, a_{\text{in}}^2}>\frac{1}{9}.
\nom}
If this condition is violated, a slow power law expansion of the Universe occurs instead of the inflation, although the energy-momentum tensor of matter reduces to the contribution of the $\La$-term.
The reason is that the contribution of the $\La$-term is quickly and fully compensated by $\ta^{\m\n}$ in the equation (\ref{6}).
We call this scenario of the Universe expansion the "compensation" scenario.

\begin{figure}[htb]
\centerline{\includegraphics[width=0.8\textwidth]{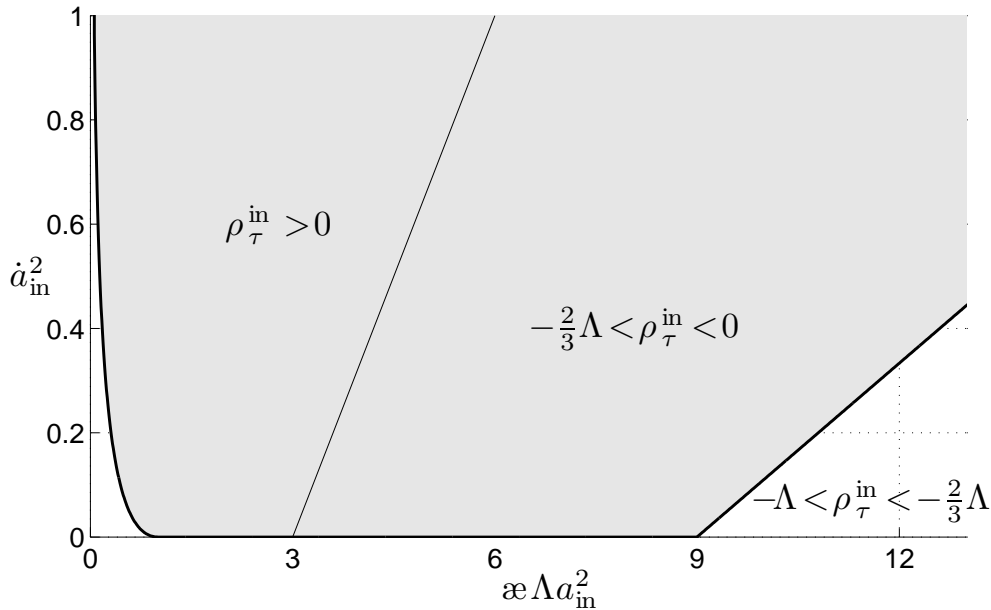}}
\caption{\label{ris3}
Domain of the initial data which lead to the inflation in case of the closed FRW model.
}
\end{figure}

For the closed FRW model ($\ep=1$) the "inflation" scenario takes place if at the same time the condition (\ref{54.1}) and at least one of the conditions (\ref{51.2}) or (\ref{51.3}) is satisfied
(the corresponding domain of initial data is shown in grey on fig.~\ref{ris3}).
In the opposite case the expansion of the Universe reaches only a certain finite value $a$ which is subject to one of two upper estimates (\ref{78}) or (\ref{85}).
We call this case the "limited" scenario.

Note that in all cases of the "inflation" scenario the value $\ro_\ta$ tends to zero at $a\to\infty$.
Using the formulas (\ref{42}) and (\ref{50}) we can easily define how do the values  $\ro_\ta$ and $p_\ta$ tend to zero in this limit:
 \disn{52}{
\ro_\ta\approx\ro_\ta^{\text{in}}\sqrt{1+\frac{\ro_\ta^{\text{in}}}{\La}}
\,\ls\frac{a_{\text{in}}}{a}\rs^4,\qquad
p_\ta\approx\frac{1}{3}\ro_\ta.
\nom}
Since just these values show the difference between the dynamics of the embedding theory in the framework of
FRW symmetry and the Einstein's dynamics (see the equations (\ref{6}) and (\ref{23})),
we can conclude that if the "inflation" scenario takes place, at the end of the inflation the deviations from the exact solutions of the Einstein equations become extremely small.
Note that for the two remaining possible scenarios -- "compensation" and "limited"{} -- it is not so.

\subsection{Expansion after inflation}
We suppose that the "inflation" scenario took place and we analyze now the expansion of the Universe at the epochs following the end of the inflation.
During the inflation the value $\ro_\ta$ decreased following the (\ref{52}), and $\ro$ changed very slowly. This is why at the end of the inflation period and immediately after it the relation $\ro_\ta\ll\ro$ was satisfied, which can be used to easily obtain from (\ref{42}) the approximate formula
 \disn{52.1}{
\ro_\ta\approx\frac{C}{\sqrt{\ro}\,a^4}.
\nom}

In order to better evaluate quantitatively the deviation of the embedding theory dynamics from the Einstein's dynamics, we introduce a dimensionless parameter
 \disn{65}{
\be=\frac{\ro_\ta}{\ro}.
\nom}
If we suppose that $(a\dot\ro_\ta)/(\ro_\ta\dot a)$ is limited by a constant (as we will see in the following, it is true for any epoch after the inflation),
then we can easily obtain from (\ref{50}) that the absolute value of $p_\ta$ cannot strongly exceed $\ro_\ta$.
This is why, if $\be\ll 1$, it means that the Einstein equations (in the framework of FRW symmetry) are satisfied with a good precision, since in the equation  (\ref{6}) the contribution $\ta^{\m\n}$ is small compared to $T^{\m\n}$.

Using the known dependencies $\ro(a)$ for the epochs where different kinds of matter play a major role, we easily to obtain, using (\ref{52.1}), the dependence of the values $\ro_\ta$ and $\be$ on $a$ for these epochs, assuming that the relation $\ro_\ta\ll\ro$ remains valid.
These dependencies read:
at the end of the inflation epoch
 \disn{66.1}{
\ro(a)\sim a^0,\qquad \ro_\ta(a)\sim a^{-4},\qquad \be(a)\sim a^{-4};
\nom}
at the ultrarelativistic matter epoch
 \disn{66.2}{
\ro(a)\sim a^{-4},\qquad \ro_\ta(a)\sim a^{-2},\qquad \be(a)\sim a^{2};
\nom}
at the non-relativistic matter epoch
 \disn{66.3}{
\ro(a)\sim a^{-3},\qquad \ro_\ta(a)\sim a^{-5/2},\qquad \be(a)\sim a^{1/2}.
\nom}
At the epoch to whose beginning belongs the current instant, and for which the main contribution to $\ro$ is given by the $\La$-term ("dark energy"), the same dependencies that for the inflation epoch are valid, i.~e. the formulas (\ref{66.1}).

It is interesting to note that at the ultrarelativistic matter epoch, as well as at the non-relativistic matter epoch, the value $\be$ showing the relative deviation from the exact solutions of the Einstein equations begins to increase.  This fact is discussed in the next section.

\section{Discussion of Universe evolution\\ in the embedding theory\label{p4}}

Now we discuss the process of the Universe expansion in the embedding theory framework, using the results of the previous section.
We adopt the usual assumption of the Big Bang theory that initially the Universe obeyed some quantum laws, but at a certain instant its characteristic size became big enough to reduce the initially important role of quantum effects and to start satisfying the classical equations of motion with a good precision. In contrast to the usual General Relativity we suppose that these equations are not the Einstein equations, but the Regge-Teitelboim ones (\ref{3}).
The state of the Universe, randomly appeared at the given instant, determines the initial conditions for the future evolution following from the classical equations.
These initial conditions are the function $y^a(x)$ defining the surface geometry, the values of the fields of matter on this surface and the time derivatives of these values, all  them given at this initial instant.

It is reasonable to suppose that at the given initial instant all the dimension characteristics of the Universe differ from the Planck values by no more than several orders of magnitude.
For example, the scaling factor of the Universe $a$ is not far from the Planck length $l_{\text{pl}}$.
It follows from this assumption that the value $\be$ introduced by the formula (\ref{65}) does not differ too much from the unity (also within several orders of magnitude),
because $\ro$, i.~e. the density of matter, and $\ro_\ta=\ta^{00}=G^{00}/\ka-\ro$, depending also on the curvature of the space, take approximately the Planck values.
Recall that $\be$ describes the relative deviation from the exact solution of the Einstein equations.

In order to solve the known problems of hot Big Bang theory, the Universe evolution must include (see for example \cite{gorbrub}) the inflation epoch with  the expansion in at least $e^{60}$ times.
In order to ensure it we first suppose as usual that at the initial instant the main role in the Universe was played by matter imitating the $\La$-term.
As shown in section~\ref{p3} assuming FRW symmetry, even in this case there may be no inflation for the embedding theory: if the "limited" scenario is implemented, the expansion is succeeded by the compression,
and the maximum size of the Universe in this case cannot strongly exceed  $l_{\text{pl}}$ (see the estimates in (\ref{78}), (\ref{85})),
and when the "compensation" scenario is implemented, the expansion is unlimited, but it follows a power law without acceleration.
Hence we introduce an additional suggestion that in our Universe the initial data allowed the "inflation" scenario at least in some region.
Also assume in this case that in this region FRW symmetry is implemented with a sufficient precision.
Note that the results of the section~\ref{p3} show that the probability of the "inflation" scenario at least is not lower than of the two others.

Assuming that during the inflation the Universe expanded at least in $e^{60}$ times, and using the (\ref{66.1}),
we can estimate $\be$ at the end of the inflation within several orders of magnitude:
\disn{66.4}{
\be\lesssim e^{-240}\approx 10^{-104},
\nom}
i.~e. it becomes very small.
But, according to (\ref{66.2}) and (\ref{66.3}), at the epochs of ultrarelativistic and non-relativistic matter $\be$ increases. Assuming that at the end of the inflation the density $\ro$ should not be higher than the Planck value, we can estimate the factor of the Universe expansion during the ultrarelativistic matter epoch as $10^{29}$.
At the same time, during the non-relativistic matter epoch, the expansion factor did not exceed $10^{4}$.
Then, using (\ref{66.2}) and (\ref{66.3}), the value $\be$ at the end of non-relativistic matter epoch can be estimated as
 \disn{66.5}{
\be\lesssim 10^{-44},
\nom}
and during the subsequent dark energy epoch the value $\be$ does not increase any more, see~(\ref{66.1}).
Thus, in spite of the possibility of growth in several epochs, the value $\be$ remains extremely small at all instants after the end of inflation.
That means (see the discussion after (\ref{65})), that during all this time the value $\ta^{\m\n}$ is only a small correction to $T^{\m\n}$,
i.~e. in the framework of FRW symmetry the Einstein equations are satisfied with a good precision.

In order to understand the precision in which the Einstein equations satisfied after the inflation in the embedding theory out of FRW symmetry framework we need to analyze the fluctuations of $\ta^{\m\n}$.
As we mentioned in the Introduction, if the Einstein's constraints are satisfied at some instant, then the Einstein equations are exactly satisfied at further evolution of the system
(technical assumptions necessary to prove this statement are given in \cite{tmf07}).
It allows to suppose with confidence
that since the Einstein equations (and in particular the Einstein's constraints) are satisfied with a very high precision (\ref{66.5}) at the instant when the deviations from FRW symmetry are small, they will also be satisfied with enough precision at all subsequent instants, when the deviations already will be not small.
Hence the fluctuations of $\ta^{\m\n}$ should the most likely be also very small.

\vskip 0.5em
{\bf Acknowledgments.}
The authors are grateful to A.~Golovnev for helpful discussions.
This work was supported in part (A.~A.~S.) by the non-profit Dynasty Foundation.

\vskip 3em

\section*{{\large Appendix:}\\ The embedding functions for all FRW models}
Following the article \cite{rozen65} we give the explicit form of embedding functions in the Minkowski space $R^{1,4}$ for all the three FRW models.

For the closed model:
 \disn{14}{
\begin{array}{lcl}
y^0=f_c(t),       &\qquad &  y^2=a(t)\sin\chi\,\cos\te,\\
y^1=a(t)\cos\chi, &\qquad &  y^3=a(t)\sin\chi\,\sin\te\,\cos\ff,\\
                  &\qquad &  y^4=a(t)\sin\chi\,\sin\te\,\sin\ff,\\
\end{array}
\nom}
where the function $f_c(t)$ is found from the condition
 \disn{15}{
\dot f_c^2-\dot a^2=1.
\nom}
The corresponding expression for the interval reads
 \disn{17}{
ds^2=dt^2-a^2(t)\ls d\chi^2+\sin^2\chi\ls d\te^2+\sin^2\te\, d\ff^2\rs\rs.
\nom}

For the open model:
 \disn{18}{
\begin{array}{ll}
y^0=a(t)\ch\chi,                    &  y^3=a(t)\sh\chi\,\sin\te\,\sin\ff,\\
y^1=a(t)\sh\chi\,\cos\te,           &  y^4=f_o(t),\\
y^2=a(t)\sh\chi\,\sin\te\,\cos\ff,  &  \\
\end{array}
\hskip -2.2em
\nom}
where the function $f_o(t)$ is found from the condition
 \disn{19}{
\dot a^2-\dot f_o^2=1.
\nom}
The corresponding expression for the interval reads
 \disn{20}{
ds^2=dt^2-a^2(t)\ls d\chi^2+\sh^2\chi\ls d\te^2+\sin^2\te\, d\ff^2\rs\rs.
\nom}

Finally, for the spatially flat model:
 \disn{21}{
\begin{array}{l}
y^0=\frac{1}{2}\ls r^2a(t)+\int\!\frac{dt}{\dot a(t)}+a(t)\rs,\\
y^1=\frac{1}{2}\ls r^2a(t)+\int\!\frac{dt}{\dot a(t)}-a(t)\rs,\\
y^2=a(t)\,r\cos\te,\\
y^3=a(t)\,r\sin\te\,\cos\ff,\\
y^4=a(t)\,r\sin\te\,\sin\ff,\\
\end{array}
\nom}
and the corresponding expression for the interval reads
 \disn{22}{
 ds^2=dt^2-a^2(t)\ls dr^2+r^2\ls d\te^2+\sin^2\te\, d\ff^2\rs\rs.
\nom}

All these embeddings are minimal, i.~e. they have a minimum possible dimension of the ambient space, equal to five in our case.
Note that if we do not limit initially the dimension of the embedding space,
then for closed and open models the assumption about the corresponding symmetry together with the Regge-Teitelboim equations lead almost unambiguously to the fact that
the surface appears to stay in a five-dimensional subspace of the ambient space $R^{1,N-1}$,
i.~e. is described by (\ref{14}) or (\ref{18})
(at some choice of the basis in the ambient space).
We show it on the example of the closed model.

Assuming that the corresponding FRW symmetry is proper to the required four-dimensional surface in the Minkowski space $R^{1,N-1}$, one can conclude that its sections $t=const$ must be three-dimensional spheres.
This conclusion seems to be the most natural, although in this case we probably reject some variants allowed by
FRW symmetry.
This is why we mentioned above that five-dimensionality of the ambient space appears {\it almost} obligatory.

It follows from this conclusion that the components of the embedding function $y^1,\ldots,y^4$ must be given by the same expressions that in the formula (\ref{14}),
and other components $y^B$ ($B=0,5,6\ldots$) must depend only on time.
Hence
 \disn{29}{
\dd_\n y^B=\de^0_\n\, \dot y^B.
\nom}
Assuming $a=B$ in (\ref{10}), using (\ref{23}) and (\ref{29}) and taking notice of $u^\m=\de^\m_0$, $g=-a^6\sin^4\chi\,\sin^2\te$ (taking into account (\ref{17})), we obtain the equation:
 \disn{31}{
\dd_0 \ls a^3\ro_\ta\dot y^B\rs=0.
\nom}
It gives
 \disn{32}{
a^3\ro_\ta\dot y^B=C^B \quad\Rightarrow\quad
\dot y^B=\frac{C^B}{a^3\ro_\ta},
\nom}
where $C^B$ is a constant vector of a subspace of the ambient space, corresponding to a part of coordinate directions $y^0,y^5,y^6,\ldots$.
One can easily see that this vector is time-like. Indeed, using (\ref{n1}) and (\ref{17}), we obtain
 \disn{33}{
(\dd_0 y^a)\eta_{ab}(\dd_0 y^b)=g_{00}=1\qquad\Rightarrow\qquad
(\dot y^B)\eta_{BB'}(\dot y^{B'})>0,
\nom}
which results, together with (\ref{32}), in the time-likeness of the $C^B$.
Since $C^B$ is a constant time-like vector, the basis of the ambient space can always be selected in order to make this vector have only a zero component.
As a result, taking (\ref{32}) into account, all components $y^B$ except $y^0$
will be constant and they can be zeroed by a shift of the ambient space. That means that the required four-dimensional surface lies in a five-dimensional subspace and hence is described by (\ref{14}).

The reasoning for the open model is quite similar.
In this case the sections $t=const$ are three-dimensional pseudo-spheres (hyperboloids) described by components of the embedding function $y^0,\ldots,y^3$ in the formula (\ref{18}).
Hence, in the closed and open FRW models the symmetry of the surface together with the Regge-Teitelboim equations give a good reason to use the embedding function (\ref{14}),(\ref{18}),
even if the dimension of the ambient space $N$ was not initially limited by five.

For the spatially flat model no similar reasoning can be used, because we cannot build for it an embedding where the constant time cross sections should be three-dimensional planes (one can easily see that it is not so for the embedding (\ref{21})).
Hence, if $N>5$, then there is probably only one reason to use in this case the embedding (\ref{21}) for the metrics corresponding to (\ref{22}), and this is the fact that
(\ref{21}) can be obtained by some limit transition from the formulas (\ref{14}) and (\ref{18}) corresponding to the open and closed FRW models.
We will obtain such a limit transition for the closed model (the reasoning for the open model is quite similar).

Substitute in (\ref{14})
 \disn{33.1}{
\chi\to \ga r,\qquad a(t)\to\frac{1}{\ga}a(t),
\nom}
and suppose that $\ga\to 0$ and the range of variation of $r$ is limited by $\ga r\ll 1$.
Then the components $y^2,y^3,y^4$ in (\ref{14}) will pass in the limit to the corresponding components of (\ref{21}). For the remaining components (having solved the equation (\ref{15}) with respect to $f_c(t)$) we obtain:
 \disn{33.2}{
y^0=\int\! dt\, \sqrt{\frac{\dot a^2}{\ga^2}+1}=\frac{a}{\ga}+\frac{\ga}{2}\int\! \frac{dt}{\dot a}+O(\ga^3),\no
y^1=\frac{a}{\ga}\cos(\ga r)=\frac{a}{\ga}-\frac{\ga}{2}r^2a+O(\ga^3).
\nom}
It can be easily shown that after the Lorentz boost with the parameter $\ga/2$ in the plane $y^0,y^1$ and the limit transition $\ga\to 0$, the components (\ref{33.2}) match (within the sign of $y^1$ which is nonessential) the corresponding components of the formula (\ref{21}).


\end{document}